\documentstyle[12pt,epsf,epsfig,wrapfig]{article}  
\begin{document}  
\newtheorem{thm}{Theorem}
\newtheorem{cor}{Corollary}
\newtheorem{Def}{Definition}
\newtheorem{lem}{Lemma}
\begin{center}  
{\large \bf  
Rotational invariance and the Pauli exclusion principle.
\ }  
\vspace{5mm}  

Paul O'Hara\footnote{Visiting Assoc. Prof, Dept of Physics, Purdue University,
W. Layayette, IN 47907, USA.}  
\\  
\vspace{5mm}  
{\small\it  
Dept. of Mathematics, Northeastern Illinois University, 5500  
North St. Louis Avenue, Chicago, IL 60625-4699, USA. email:
pohara@neiu.edu \\}  
\end{center}  
%
%
%
\begin{abstract}
In this article, the rotational invariance of
entangled quantum states is investigated as a possible  cause of the
Pauli exclusion principle. First, it is shown that a certain
class of rotationally invariant
states can only occur in pairs. This will be referred to as the
coupling principle. This in turn suggests a natural classification of
quantum systems into those containing coupled states and those that do not.
Surprisingly, it would seem that  Fermi-Dirac statistics follows
as a consequence of this coupling while the Bose-Einstein follows by breaking
it. Finally, the experimental evidence to justify the above classification
will be discussed.
Pacs: 3.65.Bz, 5.30 d, 12.40.Ee   
\end{abstract}

\section {Introduction}

Rotational invariance in quantum mechanics is usually associated 
with spin-singlet states.
In this article, after having first established a uniqueness theorem relating
rotational invariance and spin-singlet states, a statistical classification
is carried out. In effect, it will be shown that, within the construct of the
proposed mathematical model,
rotationally invariant quantum states can only occur in pairs. These 
pairs will be referred to as isotropically spin-correlated states (ISC) and
will be defined more precisely later on. This in turn will suggest a statistical
classification procedure into systems containing paired states  and those that  
do not. It will be shown that a system of
$n$ coupled and indistinguisable states obey the Fermi-Dirac statistic,
while Bose-Einstein statistics will follow when the coupling is broken.

Hopefully, the above results will help deepen our understanding
of the spin-statistics theorem first
enunciated by Pauli in his 1940 paper: ``The Connection Between Spin and
Statistics'' \cite{pauli}.
At the core of these  results is the concept of entanglement and the
work of Bell, both of which would appear to suggest that in the case of 
spin-singlet states,
microscopic causality might be violated precisely because of the
nature of entanglement and non-locality. This will be discussed in more detail 
towards the end of the article (section 7).

Throughout the paper the following notation will be used for spin
systems:\\ $\theta$ will represent a polar
angle lying within a plane such that 
$0\le \theta < 2\pi$. Denote $|\theta_j-\theta_i|$ by
$\theta_{ij}$ and write $a.e.\ \theta$ for ``$\theta$ almost
everywhere''.\\
$|\psi_{1\dots n}(\lambda_1,\dots ,\lambda_n)>$ will represent an
n-particle state, where $1\dots n$ represent particles and
$\lambda_1 \dots
\lambda_n$ represent the corresponding states.  However, if there
is no 
ambiguity oftentimes this state will be written in the more compact
form\\ 
$|\psi(\lambda_1, \dots ,\lambda_n)>\ {\rm\ or\ more\ simply\ as}\ |\psi>.$\\
$s_n(\theta)$   
will represent the spin states of particle $n$ measured in 
direction $\theta$ where $s_n(\theta)=|\pm>$.
In the case of $\theta =0$,  replace $s_n(0)$ with $s_n$ or by
$|+>$ or $|->$ according to
the context, where $+$ and $-$ represent spin up and spin down
respectively. Also let $s^-_n(\theta)$ denote the spin state
ORTHOGONAL to $s_n(\theta)$.\\ 
The wedge product of $n$ 1-forms is given by: $a_1\wedge \dots \wedge a_n=
\frac{1}{n!}\delta^{1\dots n}_{i_1\dots i_r}a^{i_1}\otimes \dots \otimes
a^{i_n}$.\\
Specifically, $a^1\wedge a^2 \wedge a^3= \frac{1}{3!}(a^1\otimes a^2 \otimes
a^3+a^2\otimes a^3 \otimes a^1 + a^3 \otimes a^1 \otimes a^2 - a^2\otimes
a^1\otimes a^3 -a^1\otimes a^3\otimes a^2 - a^3\otimes a^2 \otimes a^1).$

\section{A Coupling Principle}

The concept of isotropically spin-correlated states (to be abbreviated as
ISC) is now introduced. This definition is motivated by the probability 
properties of rotational invariance.
Intuitively, $n$ particles are isotropically spin-correlated, if 
a measurement made in an ARBITARY direction $\theta$ on ONE  
of the particles allows us to predict with certainty, the spin value of each
of the other $n-1$ particles for the same direction $\theta$.
\begin{Def} Let $H_1\otimes H_2$ be a tensor product of two
2-dimensional inner product spaces. Then $|\psi>\in H_1\otimes H_2$
is said to be rotationally invariant if
$$(R_1(\theta),R_2(\theta))|\psi>=|\psi>,$$
where each 
\begin{displaymath}
R_i(\theta)= 
\left[\begin{array}{cc}
\ \cos(c\theta) & \sin(c\theta)\\
-\sin(c\theta) & \cos(c\theta)
\end{array}\right]
\end{displaymath} 
represents a rotation on the space $H_i$ and $c$ is a constant.\cite{green}
\end{Def}
\begin{Def} Let $H_1, \dots ,H_n$ represent $n$ 
2-dimensional inner product spaces. $n$ particles are said to be isotropically 
spin correlated (ISC) if\\
(1) for all $\theta$ the two state $|\psi_{ij}>\in H_i\otimes H_j$ is 
rotationally invariant for all $i,j$ where $i\neq j$ and $1\le i,j \le n$,\\
(2) for all $\theta$ and each $m\le n$ the  state 
$|\psi>\in H_1\otimes \dots \otimes H_m$
can be written as
\begin{eqnarray}
|\psi>=\frac{1}{\sqrt2}[s_1(\theta)\otimes s_2(\theta) \dots \otimes
s_m(\theta)\pm
s^{-}_1(\theta)s^{-}_2(\theta)\dots \otimes s^{-}_m(\theta)]
\end{eqnarray}
\end{Def}
Note that it follows from the definition of ISC states that rotationally
invariant states of the form
\begin{eqnarray}
|\psi>=\frac{1}{2}(|+>|+> + |->|-> + |+>|-> - |->|+>)
\end{eqnarray}
are excluded. In other words,  the existence of ISC states means that
if we measure the spin state $s_1$ then we have simultaneously measured
the spin state for $s_2 \dots s_n$ (see Lemma 1). It can also be shown by
means of projection operators that the state defined by equation (1) is the
{\it only} state that can be projected onto the state 
$|\psi_{ij}>\in H_i\otimes H_j$ for each $i, j$. This further highlightes
its significance. 

Two examples of ISC states can be immediately given:
\begin{eqnarray} |\psi>=\frac{1}{\sqrt 2}(|+>|+>+|->|->) \end{eqnarray}
and
\begin{eqnarray} |\psi>=\frac{1}{\sqrt 2}(|+>|->-|->|+>). \end{eqnarray}
However, what is not apparent is that these are the only ISC states permitted
for a system of n-particles. This is now proven, after having first introduced
some probability notation.

Recall that $s_n(0)=|\pm>$. Associated with these states,  define
the random variable $S_n(\theta):|\pm> \to \pm 1$ such that 
$P(S_n(\theta)=1)=P(S_n(\theta)=1|S_n(0)=1)=\cos^2(c\theta)$ 
and $P(S_n(\theta)=-1)=P(S_n(\theta)=-1|S_n(0)=1)=\sin^2(c\theta)$ where $c$ is constant. 
In other words, $P$ is the spectral measure
associated with the state $|\psi(\theta)>=\cos(c\theta)|+>+\sin(c\theta)|->$. 
Similarily, a joint measure $P$ can be associated with the state
\begin{eqnarray*}|\psi(\theta_1, \theta_2)>&=&\cos(c\theta_1)\cos(c\theta_2)|+>
|+>+
\cos(c\theta_1)\sin(c\theta_2)|+>|->\\
&& + \sin(c\theta_1)\cos(c\theta_2)|->|+>
+\sin(c\theta_1)\sin(c\theta_2)|->|->\end{eqnarray*}
such that
$P(S_1(\theta_1)=+, S_2(\theta_2)=+)=P(S_1(\theta_1)=-, S_2(\theta_2)=-)=
\cos^2(c\theta_1)\cos^2(c\theta_2)$ etc.
In particular, note that $c\theta_1=\frac{\pi}{4}$ and $\theta_2=\theta$
implies $P(S_1=+, S_2=+)=\frac12\cos^2(c\theta)$.
\begin{lem} Let $|\psi>$ represent an ISC state, $|\psi_1>=
1/{\sqrt 2}(|+>+|->)$
be the initial
state of particle 1  and $S_n$ be a random variable
as defined above. If a measure $P$ on $\{S_1, \dots ,S_n\}$
exists then
$P(S_n=\pm|S_1=\pm)=1$ or $P(S_n=\pm|S_1=\mp)=1$. 
\end{lem}
{\bf Proof:} From the definition of $|\psi_1>$, it follows that
$P(S_1=+)=P(S_1=-)=\frac12$. Without loss of generality we  assume
$(S_1=+, S_2=+, \dots , S_n=+)$. But from the definition of ISC states
$P(S_1=+, S_2=+, \dots ,S_n=+)=\frac12$. Also
\begin{eqnarray*} P(S_1=+, S_2=+, \dots , S_n=+)&=&P(S_1=+)P(S_2=+, \dots ,
S_n=+|S_1=+)\\
&=&\frac12P(S_2=+, \dots ,
S_n=+|S_1=+)\\
&=&\frac12
\end{eqnarray*}
Hence, $P(S_n=+|S_1=+)=P(S_2=+, \dots , S_n=+|S_1=+)=1$.\\
\vskip 5pt
\noindent
{\bf Remark:} In terms of the physics this means that once the spin-state of  
particle 1 is measured along an axis then the values of each of the other 
particle states are
also immediately known along the same axis.
Secondly, associated with the ISC state as given in (3),
random variables $S_1$ and $S_2(\theta)$ and an associated measure $P$
can be defined such that
$P(S_1=+, S_2(\theta)=-)=(1/2)\sin^2(\theta)$ where $\theta$ can be considered
as the angle between $s_1$ and $s_2(\theta)$. 

It is now shown that it is impossible to have three or more particles in such 
ISC states. Specifically, we ask if there can exist $n$ or more ISC particles
which generate a spectral measure $P$ associated with the random 
variables $S_1(\theta_i), S_2(\theta_j), \dots S_n(\theta_k)$ and the state
$$|\psi>=\frac{1}{\sqrt 2}(s_1\otimes \dots \otimes s_n - s^-_1\otimes \dots
\otimes s^-_n)$$
on the sample space 
$H_1\otimes H_2 \dots \otimes H_n$? However, since the existence of $n$
ISC particles, presupposes the existence of $n-1$ such particles, it is
sufficient to prove the theorem for $n=3$.  As noted above, such measures 
clearly exist for $n=2$.

\begin{wrapfigure}{r}{8cm}  
\epsfig{figure=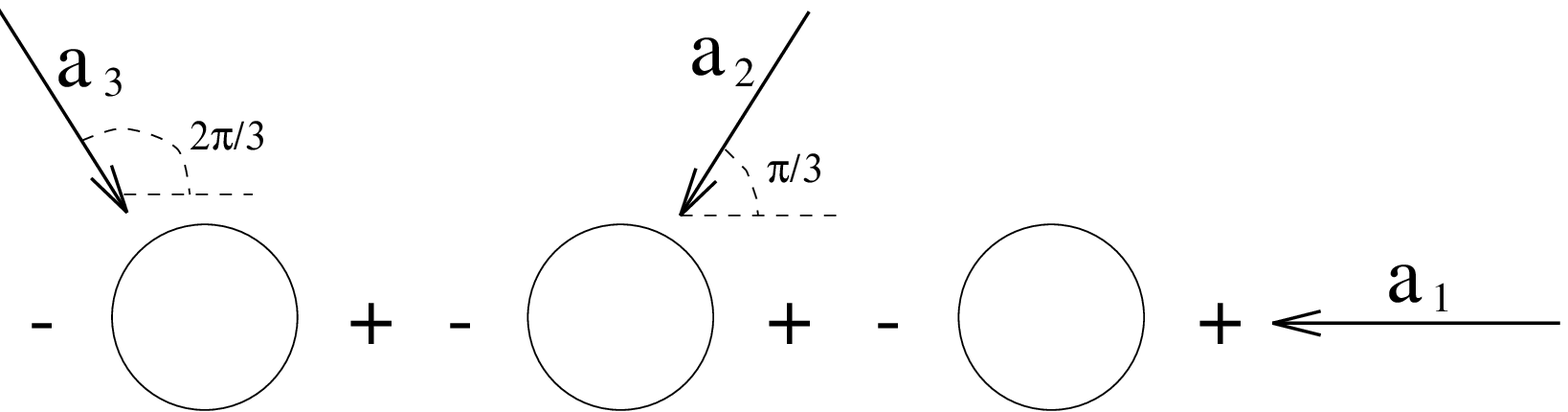,width=8cm}  
{\small Figure 1: Three isotropically spin-correlated
particles.}
\end{wrapfigure}  

In the interest of clarity, assume without loss of
generality, that there are three ISC particles such that
$s_1(\theta_i)=s_2(\theta_i)=s_3(\theta_i)$, for an arbitrary
direction $\theta_i$. This means that the joint spin state for any two of them 
is given by equation (3). 
Also, we further refine notation to write, 
for example,
$(+, +, -)$  for $S_1(\theta_i)=+, S_2(\theta_j)=+,
S_3(\theta_k)=-.$ These can also be interpreted as measurements
in the $i, j, k$ direction respectively.
Also, for the purpose of the argument
below, let $c=1/2$.
However, the argument can be made to work for any constant
value of $c$, and in a particular can be applied to the spin
of a photon, provided $c=1$. 

With notation now in place, an argument of Wigner
\cite{wig} is now adopted to
show that isotropically spin-correlated particles must occur in
pairs.  The proof is by contradiction. Specifically, consider
three isotropically spin-correlated
particles (see Fig. 1), as explained above.  It follows from the 
probability 1 condition, that three  
spin measurements can be  
performed, in principle, on the three particle system, in the  
directions  $\theta_i,\ \theta_j, \ \theta_k$. In the above
notation, this
means that $(S_1(\theta_i), S_2(\theta_j), S_3(\theta_k))$ represent the 
observed
spin values in the three different directions. Recall  
that $S_n = \pm $    
which means that there exists only two possible values for each  
measurement. Hence,for three measurements there are a total of 8 
possibilities in total.  
In particular,  
  
\vspace{-6mm}  
  
\begin{eqnarray*}  
\;\;\;\;\;\;\{(+,+,-), (+,-,-)\} & \subset &  
\{(+,+,-),(+,-,-),(-,+,-),(+,-  
,+)\}\\  
\Rightarrow \;\;\; P\{(+,+,-),(+,-,-)\} & \le &  
P\{(+,+,-),(+,-,-),(-,+,-),(+,-,+)\}.  
\end{eqnarray*}  
Therefore,  
$$\frac 12 \sin^2 \frac {\theta_{ki}}{2}\le \frac 12 \sin^2 \frac
{\theta_{jk}}{2} + \frac 12 \sin^2 \frac {\theta_{ij}}{2}$$  which
is Bell's inequality. Taking
$\theta_{ij}=\theta_{jk}=\frac {\pi}{3}$ and  
$\theta_{ki}=\frac {2\pi}{3}$ gives $\frac 12 \ge \frac
34$ which is clearly a contradiction. In other words, three  
particles cannot all be in the same spin state with probability  
1, or to put it another way, isotropically spin-correlated
particles with respect to the measure $P$ must occur in pairs.\\
\vskip 5pt
\noindent
{\bf Remark:} (1) If the ISC particles are such that 
$(s_1(\theta_i)=+,s_2(\theta_i)= -, s_3(\theta_i)=+)$ (note same
$i$) then regardless of distinguisability or not,
the spin measurements in the three different
directions $\theta_i$, $\theta_j$, $\theta_k$i can be written as: 
$$
\{(+,+,-), (+,-,-)\} \subset  
\{(+,+,-),(+,-,-),(-,-,-),(+,+  
,+)\}$$
The previous argument can now be repeated as above.  

(2) Each of the previous arguments apply also to spin 1
particles, like the photons, provided full angle formulae are
used instead of the half-angled formulae. 

\section{Pauli exclusion principle}

The above results can be cast into the form of a theorem (already
proven above) which
will be referred to as the ``coupling principle''.
\begin{thm}(The Coupling Principle) Isotropically
spin-
correlated particles must occur in PAIRS.\end{thm} 

It follows from the coupling principle that multi-particle systems can be 
divided into two categories, those containing indistinguisable coupled 
particles and
those containing indistinguisable decoupled particles. It now remains
to show that a statistical analysis of these two 
categories
generates the Fermi-Dirac and Bose-Einstein statistics respectively.

First note that in the case of ISC particles the two rotationally invariant
states can be identified with each other by identifying a spin measurement 
of $\pm$ on the second particle in
the $x$ direction with a spin measurement of $\mp$ in
the $-x$ direction, in such a way as to maintain the rotational invariance. 
In other words, by replacing the second spinor with its spinor conjugate 
\cite{car}, the state 
$$|\psi>=\frac{1}{\sqrt 2}(|+>|+>+|->|->)$$
can be written as
$$|\psi(\pi)>=\frac{1}{\sqrt 2}(|+>|->_{\pi}-|->|+>_{\pi})$$
where the $\pi$ subscript refers to the fact that the measurements on
particles 1 and 2 are made in opposite directions, while maintaining the 
rotational invariance \cite{car}.  
The state $|\psi(\pi)>$ shall be referred to as an
{\bf improper} singlet state. Furthemore, without loss of generality
the above identification means
it is sufficient to confine oneselves to singlet states when 
discussing the properties of ISC particles.
\footnote{Whether or not improper singlet states actually
exist in nature, remains an open question. 
If such states were actually discovered then they would throw new light on the 
``handedness'' problem
and their existence might possibly  be linked to parity violation.}  
It remains to show that the requirement of rotational invariance for ISC
particles generates a Fermi-Dirac type statistic. Howevever, before doing so,
it is important to emphasize that the Pauli principle has not being assumed 
but rather is being derived from the  usual form of the principle 
(written in terms of the Slater determinant) by
imposing orbital restrictions on the ISC states. 

The essential ideas are as follows:
The existence of ISC particles means that from the coupling principle (above), 
the wave function can be written uniquely
as a singlet state on the $H_1\otimes H_2$ space. It then follows by
imposing the additional requirement that ISC particles occur only
in the same orbital, that the usual singlet-state form of the wave
function can be extended to the space ${\cal S}_1\otimes {\cal S}_2$ where
${\cal S}_i=L^2({\cal R}^3)\otimes H_i$. 
Indeed, it would almost seem to be a tautology stemming
from the definition of rotational invariance. 
Nevertheless, the manner in which the notion of ISC states extend to
these spaces needs to be clarified, since
it permits an extension of the results to the usual 
$L^2 \otimes H$ space
associated with quantum mechanics and not just the more restricted spin
spaces. Once this extension is made, proof by induction can be used to derive 
the usual form of the Pauli principle 
associated with the Slater determinant, for an n-particle system.
Moreover, it is also worth noting that the existence of
spin-singlet states in general, and not only in the same orbital, permits 
more general forms of the exclusion principle (see for example Lemma 3 and
Theorem 3).
 
In what remains, let $s_n=s_n(\theta)$ represent the
spin of particle $n$ in the direction $\theta$. However, in general 
the $\theta$ can be dropped when there is no ambiguity but its presence should
always be understood.  Also, let $\lambda_n=(q_n, s_n)$
represent the
quantum coordinates of particle $n$, with $s_n$ referring to the
spin coordinate in the direction $\theta$ and $q_n$ representing
all other coordinates. In practice, 
$\lambda_n=(q_n, s_n)$ will represent the coordinates of the particle
in the state $\psi(\lambda_n)$ defined on the Hilbert space
${\cal S}_n= L^2({\cal R}^3)\otimes H_n$,
where $H_n$ represents a two-dimensional spin space of the
particle $n$.  We will mainly work with $\lambda_n$ but sometimes,
there will be need to distinguish the $q_n$ from the $s_n$. This
distinction will also allow the ket to be written as:
$|\psi(\lambda)>=|\psi(q)>s=|\psi(q)>\otimes s$ where $s$ represents 
the spinor.
With these distinctions made, the notion of orbital is now defined 
and a sufficient condition for obtaining the usual form of the 
Fermi-Dirac statistics within the context of our mathematical model, is given. 
\begin{Def}Two particles whose states are given by $|\psi(q_1,s_1)>$ and 
$|\psi(q_2,s_2)>$ respectively are said to be in the same q-orbital when
$q_1=q_2$.
\end{Def}
The following Lemma allows us to extend the results for ISC particles
defined on the space $H_1\otimes H_2$ to the larger space 
${\cal S}_1\otimes {\cal S}_2$.
As mentioned above,  the Pauli 
principle is not being assumed but rather is being deduced by invoking
rotational invariance of the ISC particles. 
Conversely, if the rotational invariance condition
is relaxed then the Pauli principle need not apply and as a result many
particles can be in the same orbital. 
\begin{lem} Let
\begin{eqnarray*}
|\psi(\lambda_1,\lambda_2)>
&=&c_1|\psi_1(\lambda_1)> \otimes
|\psi_2(\lambda_2)>+c_2|\psi_1(\lambda_2)> \otimes
|\psi_2(\lambda_1)>,\end{eqnarray*} where $c_1, c_2$ are independent of
$\lambda_1$, $\lambda_2$, represent an indistinguisable two particle
system defined on the space ${\cal S}_1\otimes {\cal S}_2$.
If ISC states for a system of two indistinguisable
and non-interacting particles occur only in the same q-orbital  
then the system
of particles can be represented by the Fermi-Dirac statistics.
\end{lem}
{\bf Proof:} The general form of the non-interacting and indistinguisable two
particle state is given by
\begin{eqnarray*}
|\psi(\lambda_1,\lambda_2)>
&=&c_1|\psi_1(\lambda_1)> \otimes
|\psi_2(\lambda_2)>+c_2|\psi_1(\lambda_2)> \otimes
|\psi_2(\lambda_1)>\\
&=&c_1|\psi_1(q_1)>s_1 \otimes
|\psi_2(q_2)>s_2+c_2|\psi_1(q_2)>s_2 \otimes
|\psi_2(q_1)>s_1\end{eqnarray*}
where $c_1,\ c_2$ are constants, such that $c^2_1=c^2_2=1/2$ for all
$\lambda_1$ and $\lambda_2$.
Let $q_1=q_2$ then the particles are in the same q-orbital and hence
rotationally invariant by the ISC condition. Therefore, $c_1=-c_2$ and 
$$|\psi(\lambda_1,\lambda_2)>=\frac{1}{\sqrt 2}[|\psi_1(\lambda_1)> \otimes
|\psi_2(\lambda_2)>-|\psi_1(\lambda_2)> \otimes
|\psi_2(\lambda_1)>].$$
The result follows. QED\\
\vskip 5pt
\noindent
{\bf Remark}: (1) The above lemma also applies to improper singlet states,
in other words to particles whose spin correlations are parallel to each other 
in each direction. This can be done by correlating a measurement in
direction $\theta$ on one particle, with a measurement in
direction $\theta + \pi$ on the other.  In this case, the state
vector for the parallel and anti-parallel measurements will be
found to be: 
$$|\psi(\lambda_1,\lambda_2)>=\frac{1}{\sqrt
2}[|\psi(\lambda_1)>
\otimes
|\psi(\lambda_2(\pi))>-|\psi(\lambda_2)> \otimes
|\psi(\lambda_1(\pi))>]$$
where the $\pi$ expression in the above arguments, refer to the
fact that the measurement on particle two is made in the opposite
sense, to that of particle one.\\
(2)If the coupling condition (rotationally invariance) is removed then a 
Bose-Einstein
statistic follows and is of the form:
$$|\psi(\lambda_1,\lambda_2)>=\frac{1}{\sqrt
2}[|\psi(\lambda_1)>
\otimes
|\psi(\lambda_2)>+|\psi(\lambda_2)>\otimes
|\psi(\lambda_1)>].$$
This will be discussed in more detail later. See, for example, Corollary 1
following Lemma 3 and Corollary 2 following Theorem 4. We now prove the 
following theorem:
\begin{thm} (The Pauli Exclusion Principle) A
sufficient condition for a system of n indistinguishable
and non-interacting particles defined on the space ${\cal S}_1\otimes \dots
{\cal S}_n$
to exhibit Fermi-Dirac statistics is that it contain spin-coupled
q-orbitals.\end{thm}
{\bf Proof:} It is sufficient to work  with three particles, but it should be
clear that the argument can be extended by induction to an n-particle
system.  Consider a system of three
indistinguishable particles, containing spin-coupled particles.
Using the above notation and applying
Lemma 2 to the coupled particles in the second line below, gives:
\begin{eqnarray*}|\psi (\lambda_1, \lambda_2,
\lambda_3)>
&=&\frac{1}{\sqrt 3}
\{|\psi(\lambda_1)>\otimes |\psi(\lambda_2,\lambda_3)> +
|\psi(\lambda_2)>\otimes |\psi(\lambda_3,\lambda_1)>\\ &\
&\hspace{5mm}
+|\psi(\lambda_3)>\otimes |\psi(\lambda_1,\lambda_2)>\}\\
&=&\frac{1}{\sqrt {3!}}\{|\psi(\lambda_1)>\otimes
[|\psi(\lambda_2)>\otimes |\psi(\lambda_3)>-
|\psi(\lambda_3)>\otimes |\psi(\lambda_2)>]\\ 
\ & &+|\psi(\lambda_2)>\otimes [|\psi(\lambda_3)>\otimes
|\psi(\lambda_1)>-
|\psi(\lambda_1)>\otimes |\psi(\lambda_3)>]\\ 
\ &&+|\psi(\lambda_3)>\otimes [|\psi(\lambda_1)>\otimes
|\psi(\lambda_2)>-
|\psi(\lambda_2)>\otimes |\psi(\lambda_1)>]\}\\
&=&\sqrt{3!}|\psi_1(\lambda_1)>\wedge |\psi_2(\lambda_2)>\wedge
|\psi_3(\lambda_3)>
\end{eqnarray*} where $\wedge $ represents the wedge product.
Thus the wave function for the three indistinguishable particles
obeys Fermi-Dirac statistics.  The n-particle case follows by
induction. QED\\
As an example, consider
the case of an ensemble of 2n identical non-interacting particles
with discrete energy levels
$E_1, E_2, \dots $, satisfying the Fermi-Dirac statistics as above then 
all occurances
of such
a gas would necessarily have a twofold degeneracy in each of the discrete 
energy levels and the lowest energy would be given by
$$E=2E_1+2E_2+2E_3+\dots +2E_n.$$  

The above theorem applies only under certain conditions. However, both in 
theory and practice,  the spin-singlet state
do not have to be confined to the same q-orbital, as for example in the
case of the 1s2s-electron configuration in He. 
This in turn requires a more general formulation of an exclusion principle:
\begin{lem} Let $|\psi (\lambda_1,\lambda_2)>\in L^2({\cal R}^3)
(q_1,q_2)\otimes H_1
\otimes H_2$ denote
the state of two indistinguisable particles where the 
$\lambda_1$ and
$\lambda_2$ are as defined above. If the particles are in a spin-singlet
state then 
$$  
|\psi(\lambda_1,\lambda_2)>= \frac{1}{\sqrt 2}[|\psi (q_1,q_2)>s_1\otimes s_2
-|\psi (q_1, q_2)>s_2\otimes s_1].$$
\end{lem}  
{\bf Proof:}
The general form for the indistinguisable two particle
state is given by
$$|\psi(\lambda_1,\lambda_2)>=c_1|\psi(q_1,q_2)>
s_1\otimes s_2 +c_2|\psi(q_2, q_1)> 
s_2 \otimes s_1.$$
Invoking rotational invariance of the spin-singlet state gives\newline
$c_1|\psi_1(q_1,q_2)> = - c_2|\psi(q_2,q_1)>$, from the linear independence
of the $s_1\otimes s_2$ states.
Normalizing the wave function gives
$|c_1|=\frac{1}{\sqrt{2}}$. The result follows. 
QED\\ 

\begin{cor} Let $|\psi (\lambda_1,\lambda_2)>\in L^2(R^3)(q_1,q_2)\otimes H_1
\otimes H_2$ denote
the state of two indistinguisable particles where the 
$\lambda_1$ and
$\lambda_2$ are as defined above, then 
$$  
|\psi(\lambda_1,\lambda_2)>= \frac{1}{\sqrt 2}[|\psi (q_1,q_2)>s_1\otimes s_2
+|\psi (q_2, q_1)>s_2\otimes s_1].$$
\end{cor}  
{\bf Proof:} The general form of the two particle state is given by
$$|\psi(\lambda_1,\lambda_2)>= c_1|\psi (q_1,q_2)>s_1\otimes s_2
+c_2|\psi (q_2, q_1)>s_2\otimes s_1.$$
Indistinguisability implies that $c^2_1=c^2_2=1/2$. Since the particles are
not necessarily in a singlet state, then it is possible to  consistently choose
$c_1=c_2=\frac{1}{\sqrt 2}$, in this case.  The result
follows. QED\\ 

Lemma 2, Lemma 3 and Corollary 1 taken together can now be used to
give a complete classification of the 1s2s state of He.  
The $SU(2)\otimes SU(2)$
properties of spin give the decomposition $2\otimes 2 = 3 \oplus 1$. 
Corollary 1 gives the triplet state composition of the 1s2s configuration. 
Lemma 3 gives the singlet state configuration in general.  
All of these states are 
experimentallay verified \cite{ham}.  Lemma 2 on the other hand,
indicates that the
Slater determinant permits ONLY the triplet state of the 1s2s
configuration and the $1s^2$ singlet. It precludes the 1s2s singlet state.
Lemma 3 defines the spin singlet state which is antisymmetric under an
exchange of spin. In contrast, Lemma 2 defines the spin singlet state
which is antisymmetric under complete particle exchange and not just spin alone.

To conclude this section, Lemma 2 and Theorem 2 express a Pauli type exclusion 
principle 
which follows naturally from the
coupling principle, or equivalently the rotational invariance. 
Either two particles are ISC and obey a Fermi-Dirac
type statistic or they are statistically independent of each other and
obey a Bose-Einstein statistic.  In the next section,  
Theorem 2 is generalized to cover interacting n-particle systems.

\section{Multi-particle interacting systems}
Attention is now turned to interacting particles. Once again,
the coupling principle is sufficient to derive an
exclusion principle in this case. The burden of the proof
rests on making full use both of the notions of isotropy and 
indistinguisability. Isotrophy, in fact, will be  introduced by working 
with the Fock spaces.

Let $F[n]=H(\theta_1)\oplus \dots \oplus H(\theta_n)$ be a
Fock space, where each $H(\theta_i)$ is a two-dimensional Hilbert space.
If $s[n]\in F[n]$ then  $s[n]$ is said to be a spin n-state. The scalar product 
of two vectors $s_1[n]$ and $s_2[n]$ in $F[n]$ is defined by
$(s_1[n],s_2[n]) = \sum_{k=1}^ns_1(\theta_k)s_2(\theta_k).$
We then say that two particles are in the same spin n-state if $s_1[n]=s_2[n]$ 
and in OPPOSITE 
spin-states if $s_1(\theta_k)=s^-_2(\theta_k)$ for each $k$. 

Specifically, consider 
n interacting particles as defined on the space\newline
$L^2({\cal R}^3)(q_1, \dots ,q_n)\otimes
F_1[m]\otimes \dots \otimes F_n[m]$, where $m\ge n$ and $F_k[m]$ represents 
the Fock spin space of the particle $k$. It should be noted that the use of the
Fock space is essential and cannot be replaced by a Hilbert
space of the form $H_1\otimes \dots \otimes H_n$,  otherwise the results 
below would  become meaningless.
For example, in the case of spin measurements on three electrons in
the same direction $\theta$, two of them must necessarily be in the same 
spin state. This means that for the above Hilbert space representation,
the only possible Fermi-Dirac
state for the electrons would be the vacuum state which is impossible. 
However, by introducing Fock spaces for the spin, not only can three electrons 
be in different spin states, but 
only two different statistical structures 
emerge naturally, namely,  those containing coupled particles and those which do not.

For example, in the case of three particles, denoting the spin state 
$s_k[m]$ by $s_k$, gives 
\begin{eqnarray*}
\sqrt 3!|\psi(\lambda_1,\lambda_2,\lambda_3)>&=&\ |\psi(q_1,q_2,q_3)>
s_1\otimes s_2\otimes s_3+|\psi(q_2,q_1,q_3)>s_2 \otimes s_1
\otimes s_3\\
&&+|\psi(q_2,q_3,q_1)>
s_2\otimes s_3\otimes s_1+|\psi(q_3,q_2,q_1)>s_3 \otimes s_2
\otimes s_1\\
&&+|\psi(q_3,q_1,q_2)>
s_3\otimes s_1\otimes s_2+|\psi(q_1,q_3,q_2)>s_1 \otimes s_3
\otimes s_2.
\end{eqnarray*}
Now imposing spin-singlet state coupling on 
$|\psi(\lambda_1, \lambda_2, \lambda_3)>$ for each pair $\lambda_i, \lambda_j$
yields 
$|\psi(q_{i_1},q_{i_2},q_{i_3})>=|\psi(q_1,q_2,q_3)>$ for
for every even permutation and\newline 
$|\psi(q_{i_1},q_{i_2},q_{i_3})>= -
|\psi(q_1,q_2,q_3)>$ for every odd permutation. Hence
\begin{eqnarray*}
\sqrt 3!|\psi(\lambda_1,\lambda_2,\lambda_3)>&=& |\psi(q_1,q_2,q_3)>
(s_1\otimes s_2\otimes s_3-s_2 \otimes s_1
\otimes s_3\\
&&\ \ +
s_2\otimes s_3\otimes s_1-s_3 \otimes s_2
\otimes s_1
+s_3\otimes s_1\otimes s_2-s_1 \otimes s_3
\otimes s_2).
\end{eqnarray*}
This can be expressed more formally with the following
theorem:
\begin{thm}In a system of n indistinguisable particles containing
a spin-singlet pair, no two particles can be in the same state
and the general form of the state function in  
$L^2({\cal R}^3)(q_1, \dots ,q_n)\otimes
F_1[m]\otimes \dots \otimes F_n[m]$, where $m\ge n$ and $F_k[m]$ is a Fock  
space, is given by:
$$|\psi(\lambda_1, \dots,\lambda_n)>=|\psi(q_1, \dots
q_n)>\sqrt{n!}s_1\wedge \dots \wedge
s_n.$$\end{thm}

{\bf Proof:}  Let $\sigma_P$ denote a permutation of the particle
states. The general form of an n particle indistinguisable
state for the system under discussion, is
given by
$$|\psi(\lambda_1, \dots,\lambda_n)>=\frac{1}{\sqrt n!}
\sum\sigma_P|\psi(q_1, \dots q_n)>
s_1\otimes\dots \otimes s_n.$$
But, indistinguisability and the rotational invariance of the
singlet coupling gives 
for every
transposition $(ij)$ 
$$|\psi(q_1,\dots q_i \dots q_j \dots , q_n)> = -|\psi(q_1, \dots
q_j \dots q_i \dots , q_n)>.$$ This means that
$\sigma_P|\psi(q_1, \dots ,q_n)>= |\psi(q_1,
\dots ,q_n)>$ for every even permutation and  
$\sigma_P|\psi(q_1, \dots ,q_n)>=
-|\psi(q_1, \dots ,q_n)>$ for every odd permutation.
It follows immediately that for a.e $\theta$ that
$$|\psi(\lambda_1, \dots,\lambda_n)>=|\psi(q_1, \dots q_n)>
\sqrt{n!}s_1\wedge \dots \wedge s_n .$$
The theorem is proven. QED

\noindent Remark:(1) In the general case, no claim is being made as to the
properties of\newline
 $|\psi(q_1, \dots ,q_n)>$. Ideally the boundary
conditions of each problem will dictate these properties. Also,
there is no reason why this state vector should be symmetric or anti-
symmetric under a change of 
coordinates nor are such conditions necessary to formulate a 
Pauli-type exclusion 
principle. In fact, our construct has yielded a Fermi-Dirac type 
statistic with respect
to the spin operator of the system, independently of the
properties
of $|\psi(q_1, \dots ,q_n)>$. However, it is always possible to impose further
restrictions in terms of energy density or locality requirements, to further 
restrict the structure.  
(2) The properties of spin-type systems help explain why in chemistry
only two electrons
share the same orbital, why all chemical bonding involves pairs of
electrons  and why Cooper pairs occur in the theory
of superconductivity.\cite{epr},\cite{spin}
\vspace{3mm}    

\section{Bose-Einstein Statistics}

In the above discussion rotational invariance has played a key role
in formulating a Fermi-Dirac 
statisticis for multi-particle ISC systems as defined by definition 1, 2 and
3. Indeed,  from the perspective of this paper, it would seem to be the 
underlying cause of
the Pauli exclusion principle.  It now remains to investigate the statistics of 
multiparticle systems when this condition is relaxed.  As noted above
the rotational invariance implies that ISC particles can be written in the
form of a singlet state, (either proper or improper).  Moreover, the definition
of indistinguisability means that there is no bias in favor of any of the
components of the permutable states. For example, if
$$|\psi(\lambda_1,\lambda_2)>=a|\lambda_1>\otimes |\lambda_2>+b|\lambda_2>
\otimes |\lambda_1>$$
is permutable then $a^2=b^2$, otherwise if $|a|>|b|$ (respectively  $|b|>|a|$) 
there 
would be a bias in favor of the state associated with $a$ (respectively $b$)
which, together with the law of large numbers,  could then be used to 
partially distinguish the states.
\begin{thm} Permutable states for a system of $n$ non-interacting particles,
defined on the space ${\cal S}_1\otimes \dots \otimes{\cal S}_n$, obey either 
the Fermi-Dirac or the
Bose-Einstein statistic.
\end{thm}
{\bf Proof:} Let 
$$\sigma|\psi(\lambda_1, \dots ,\lambda_n)>=\sum^{n!}_{i=1}c_i
|\psi(\sigma_i \lambda_1)>\otimes \dots 
\otimes |\psi(\sigma_i \lambda_n)>$$
where $\sigma_i$ represents a permutation of the particles in the states
$\lambda_1, \dots , \lambda_n$. We now claim that if the system of 
indistinguisable particles are not in the Fermi-Dirac state then
$$\sigma(|\psi(\lambda_1, \dots ,\lambda_n)>=\frac{1}{\sqrt n!}\sum^{n!}_{i=1}
|\psi(\sigma_i \lambda_1>\otimes \dots 
\otimes|\psi(\sigma_i\lambda_n)>.$$
This follows by noting that if $c_i=\frac{1}{\sqrt n!}$ and 
$c_{i+1}=-\frac{1}{\sqrt n!}$
for each $i$ then Fermi-Dirac statistics results. Hence, assume that there
is not an exact pairing and that $c_1=\dots c_i=\frac{1}{\sqrt n!}$ 
where either
$i>\frac{n!}{2}$ or $i< \frac{n!}{2}$
and $c_{i+1}=\dots =c_{n!}=-\frac{1}{\sqrt n!}$.
Then taking $\lambda_1=\lambda_2$ many of the terms on the right hand 
side (in fact $\min\{2i, 2(n!-i)\}$ terms) will cancel, leaving only the excess unpaired
positive (negative) terms.  If the remaining number of terms in the
expansion is less than $n!$ then $|\psi(\lambda_1,\dots,\lambda_n)>$ is
NOT invariant under the complete set of permutations, which is a 
contradiction. It follows that the
number of terms must be $n!$ and nothing vanishes. Hence
$|\psi(\lambda_1,\dots ,\lambda_n)>$ exhibits Bose-Einstein statistics.
The result follows.
\vskip 5pt
\noindent
It might be instructive to apply the above theorem to a three particle wave
function that is not of the above type. Consider:
\begin{eqnarray*}|\psi (\lambda_1, \lambda_2,
\lambda_3)>
&=&\frac{1}{\sqrt {3!}}\{|\psi(\lambda_1)>\otimes
[|\psi(\lambda_2)>\otimes |\psi(\lambda_3)>+
|\psi(\lambda_3)>\otimes |\psi(\lambda_2)>]\\ 
\ & &+|\psi(\lambda_2)>\otimes [|\psi(\lambda_3)>\otimes
|\psi(\lambda_1)>+
|\psi(\lambda_1)>\otimes |\psi(\lambda_3)>]\\ 
\ &&+|\psi(\lambda_3)>\otimes [|\psi(\lambda_1)>\otimes
|\psi(\lambda_2)>-
|\psi(\lambda_2)>\otimes |\psi(\lambda_1)>]\}.
\end{eqnarray*} 
On putting $\lambda_1=\lambda_2$,
\begin{eqnarray*}|\psi (\lambda_1, \lambda_2,
\lambda_3)>
&=&\frac{1}{\sqrt {3!}}\{|\psi(\lambda_1)>\otimes
[|\psi(\lambda_2)>\otimes |\psi(\lambda_3)>+
|\psi(\lambda_3)>\otimes |\psi(\lambda_2)>]\\ 
\ & &+|\psi(\lambda_2)>\otimes [|\psi(\lambda_3)>\otimes
|\psi(\lambda_1)>+
|\psi(\lambda_1)>\otimes |\psi(\lambda_3)>] 
\end{eqnarray*}
which is not invariant under permutations.

\begin{cor} Let
\begin{eqnarray*}|\psi(\lambda_1,\dots \lambda_n)>
&=&\sum^{n!}_1 c_i|\psi(\sigma \lambda_1)> \otimes \dots
|\psi(\sigma \lambda_n)>,\end{eqnarray*}
where $c_i$ are independent of
$\lambda_i$, represent an indistinguisable $n$-particle
system defined on the space ${\cal S}_1\otimes \dots \otimes {\cal S}_n$.
This  system
of particles can be represented by the Bose-Einstein statistics.
\end{cor}
{\bf Proof:} Normalizing the wave function and using indistinguisability
gives $c^2_i=c^2_j$, for each $i$ and $j$.  If $c_i=-c_{i+1}$
then each q-orbital
would be a spin-singlet
state.  But this is not so.  Hence $c_i=c_j$ by the previous lemma 
and the result follows.
 
Denote the set of permutations that leave invariant the Bose-Einstein and 
Fermi-Dirac
statistics by $s_n$ and $a_n$, respectively. It follows that certain types
of mixed statistics can be now described.  For example, the 2-electrons
of a helium atom, considered together with the 3-electrons in a lithium 
atom obey $a_2\otimes a_3$ statistics while the 5 electrons in the boron atom
obey $a_5$ Fermi-Dirac statistics.  The electrons in three
different helium atoms obey $a_2\otimes a_2 \otimes a_2$ if the helium atoms
are considered distinguisable and $s_3\circ (a_2\otimes a_2\otimes a_2)$
if the atoms are indistinguisable.
Finally, if we consider collectively the $n$ distinct electrons in $n$
indistiguisable hydrogen atoms then these $n$ electrons can be
described with $s_n$ Bose-Einstein statistics.

\section {Clarification by Contrast}

It now remains to discuss the above mathematical results from the perspective
of Pauli's famous paper on spin-statistics \cite{pauli} and in the overall 
context of the experimental evidence (discussed next section). 

First, in Pauli's construct the principle of microscopic causality 
($[\Theta(x), \Theta(y)]=0$ for $(x-y)^2<0$ where $x$ and $y$
represent Minkowski 4-vectors) 
underlies the distinction between bosons and fermions.
In contrast, rotationally invariant states (particles) which are key to 
understanding the model presented,
would seem to violate this principle by definition.  
Specifically, in Pauli's formulation if $\sigma_i(x)$, $\sigma_j(y)$
represent the spin operators for particles located at $x$ and $y$ respectively
then particles which are statistically independent of each other can have
their joint spin operators represented by 
$I\otimes \sigma_i(x)$ and $\sigma_j(y)\otimes I$ respectively. This reflects
the fact that the joint system represented by the Hilbert spaces $H_1
\oplus H_2$ can be decomposed into two coherent systems $H_1$
and $H_2$, and the joint state for the two particles is a mixture of
states from $H_1$ and $H_2$.
Moreover, since these commute both for timelike and spacelike separations then
microscopic causality is obeyed.  However, in the case of two particles in a 
spin-singlet state no such decomposition is possible.  The spin-singlet 
state is a pure state and obeys the principle of superposition, even at
spacelike separations. Hence, it
cannot be written as a mixture of two states governed by superselection
rules. Moreover, a measurement made on one particle of the singlet state
means that we have SIMULTANEOUSLY measured the spin of the other
particle in the same direction (see Lemma 1), but it does not permit us
to make a second independent measurement in a different direction,
without destroying the predicted value for the second particle.
In other words,
for pure singlet states $[\sigma_i(x), \sigma_i(y)]=0$ for all $x$ and $y$
but $[\sigma_i(x), \sigma_j(y)]\neq 0$ for all $x$ and $y$, $i\neq j$, including
spacelike intervals.  Therefore, it would seem that microscopic causality is 
violated for entangled states and
the existence of such states constitute non-local events.  However, it should 
be emphasized that non-locality does NOT mean a
breakdown of cause and effect (at least in this case); it simply means that  
``alla Pauli" we are dealing with the
mathematics of non-commuting operators and the statistics of predetermined
correlations which remain correlated even when the distance between the
particles is spacelike. Any other usage of the word ``non-locality'' 
is NOT intended.

Secondly, if the principle of microscopic causality were to be violated
then as a consequence particles in themselves would be neither fermions 
nor bosons,
but rather the relationship between the particles would determine whether
the multi-particle system obeyed Fermi-Dirac or Bose-Einstein statistics.
In our construct, indistinguisable ISC particles behave as fermions(Theorem 2), 
but once the same particles are disentangled
they behave as bosons (Cor. 2). For this reason, as already noted previously,
the consequences of coupling can be applied also to spin 1 particles like
photons.  In our formulation, singlet-state photons as used in the Aspect et al.
experiment \cite{asp}, 
are an (albeit trivial) instance of Fermi-Dirac statistics.

The above classification is also consistent with Pauli's own work. 
He is very aware that if he includes the possibility of non-local
events in his schema then his demand that particles with integral spin
not obey an exclusion principle would fall apart. 
To quote him:
`` For integral spin the quantization
according to the exclusion principle is not possible. For this result it is
essential the use of the $D_1$ function in place of the $D$ function be, for
general reasons, discarded.'' 
In fact, it is precisely the class of   
$D_1(x,x_0)$ solutions of the second order wave equation (\cite{pauli} p 720), 
that underlies  non-local events. In his own indirect way, Pauli is indeed
affirming that non-locality permits bosons to be second quantized as fermions.

Thirdly, as pointed out above,  in Pauli's original discussion 
the distinction between entangled states and non-entangled states does not
arise. This is understandable, since the significance of entanglement only
emerged in the 1960's with the work of Bell and others. This also has 
consequences for the theory of angular momentum. Specifically,
Pauli assumes that particle wave functions obey the rule 
$U_1(j_1,k_1)U_2(j_2,k_2)=\sum_{j,k}U(j,k)$ (\cite{pauli}, p717) where
$j=j_1+j_2, j_1+j_2-1, \dots, |j_1-j_2|$ and $k=k_1+k_2, k_1+k_2-1, \dots ,
|k_1-k_2|$. In fact, it would appear that this rule only applies to 
disentangled particles.
As a counter example, consider two spin-1/2 electrons in
a singlet state.  The above decomposition, if it were valid, would mean
that $U(1/2)U(1/2)=U(1)\oplus U(0)$ but we know that for a singlet state
only the $U(0)$ case arises.  This can be seen more clearly, if written
in terms of probability theory. Let $S_1, S_2$ be two independent and 
identically ditributed random variables 
such that for $i=1$ or $i=2$, $P(S_i=1/2)=P(S_i=-1/2)=1/2$. Let $M(S_1)$ and
$M(S_2)$ be the probability moment generating functions of $S_1$ and $S_2$
respectively, then it is easy to show that $M(S_1)M(S_2)=M(S_1+S_2)$. However,
once the independence condition is relaxed the above multiplication rule is
no longer valid.

\section{Experimental evidence}

The experimental justification for accepting the new form of the
spin-statistics theorem would appear to come from a wide range of physical
phenomena. First, note that the existence of photons in the spin-singlet
state seems to support the above formulation.
Secondly, we will argue that the new approach offers a more unified and 
coherent explanation of the phenomenon of paramagnetism. Thirdly, the existence 
of Cooper pairs in superconductivity can be explained as a specific instance of
ISC particles. Fourthly, we discuss baryonic
structure from the new perspective. Finally, in keeping with
the tradition of theoretical physics, a  prediction will be made about the
probability distribution for the spin decomposition of a beam of ionized
deuterons,  a prediction which will distinguish it from the current theory.

(1)  Rotational invariance demands  the wave function for spin-singlet-state 
photons to be of the form
$$|\psi> = \frac{1}{\sqrt 2}(|+>|->-|->|+>).$$ By definition this is 
a Fermi-Dirac type statistic.  Spin-singlet-state photons were at the heart of 
the  Aspect experiment \cite{asp} 
and hence their existence has already been verified.
The exclusion principle is then a tautology in the sense that while
photons are in a spin-singlet state then both of them cannot be in the
same state.
Note, however, that the fermionic state of photons can easily be destroyed 
by experiment and forced into a Bose-Einstein state. It is a trite (but
nevertheless valid) application of the exclusion principle.

(2) The theory of paramagnetism yields two different equations
for the magnetic susceptibility, one given by the classical Langevin (Curie)
function which makes no reference to the Pauli exclusion principle and the
other which is derived as a direct application of the exclusion 
principle. It would appear that our formulation of the exclusion principle
gives an equally apt understanding of the phenomenam and would 
appear to further clarify Pauli's explanation, by focusing on the unique
role of the non-spin-singlet states. 
Specifically, when the magnetic field is turned on, 
the spin up component of the spin-singlet state has its energy 
shifted down by $\mu B$ while the spin down component has its energy
shifted up by $\mu B$, with the spins being aligned 
into  parallel and anti-parallel states, resulting in a net contribution to
the magnetic field of 0. 
Hence, the paired elecrons contribute nothing to the magnetic 
susceptibility.  The remaining unpaired electrons act in such a way that 
there is an excess of electrons in the spin up state over the spin down
state, in order to maintain the common electrochemical potential. Specifically, 
if we  let $g(\epsilon)$ be the electron density of
available states per unit energy range then the total excess energy is given
by $g(\epsilon_F)\mu B$, provided $\mu B<<\epsilon_F$, which is
Pauli's result for paramagnetism.  It should also be pointed out that 
from the perspective of Pauli's version of the spin-statistics theorem, 
half-integral-spin particles such as electrons  or 
gaseous-nitric-oxide (NO) molecules, remain as fermions regardless of  
thermodynamic considerations or of the state they occupy.
However, it is generally taken forgranted 
\cite{hall} that as $kT>>\epsilon_f$ the Pauli principle no longer applies 
and the magnetic susceptibility is in this case best estimated by using the
Boltzmann statistics.  This gives rise to the ambivelant situation of referring
to particles as fermions, although they are no longer subjected to the Pauli  
exclusion principle.  With our approach, this ambiguity is removed and  
a more natural and coherent explanation of the transition from Fermi-Dirac to
Boltzmann statistics is forthcoming. Essentially,
the Boltzmann statistics emerges when the spin coupling which seems to be the 
underlying causes of the Fermi-Dirac statistics, is first broken
and the particles  then move apart to become distinguisable and 
statistically independent.
This breaking of the coupling occurs naturally when the 
temperature is raised, and they become distinguisable and independent 
when the distance separating the particles become large enough to overcome
interactions between the particles. As a result, the particles obey Boltzmann
statistics and 
``the Curie law applies to paramagnetic atoms in a low density gas, just
as to well separated ions in a solid...".\cite{hall}.

(3) The existence of Cooper pairs as spin-singlet states in the theory of
superconductors is another instance of the coupling principle at work. 
Moreover, the fact that $2n$ superconducting electrons exhibit the statistics 
of $n$ boson pairs and NOT the usual $a_{2n}$ Fermi-Dirac statistics
(\cite{hall}, p268), normally associated with the exclusion principle, 
again suggests that the
current definition of bosons and fermions in terms of quantum number is 
inadequate. In contrast, this paper classifies particles into coupled or
decoupled particles and then permits various statistics to emerge in
accordance with the degree of indistinguishability that is imposed on the 
system. When complete indistinguisability is imposed on the system then
Fermi-Dirac or Bose-einstein statistics will ensue according as to whether
the system permits coupled (Theorem 2) or only decoupled particles (Theorem 4)
respectively. On the other hand, if complete indistinguisability is relaxed in
favor of some type of partial indistinguisability (as with Cooper pairs),
we obtain different types of mixed statistics (see for example the last
paragraph of section 5).

(4) Spin $\frac 32$ baryons may be viewed as
excited states of spin $\frac 12$ baryons.  In particular, spin $\frac 32$
quarks may be viewed as composed of three quarks with uncorrelated spin
states (statistically independent) while the spin $\frac 12$ baryon would
contain a pair of quarks in a singlet state. Moreover, the need for color
to explain the structure of $\Delta^{++}$ and $\Omega^-$ particle, becomes
unnecessary in the new approach. Of course, this does not preclude the use
of color to give ``colorless'' baryons. \cite{qui}
   
(5) It is well known that the deuteron ion is in a spin-triplet
state. Denote the possible observed spin values $X$ by +1, 0, -1 respectively.  
Conventional quantum mechanics
predicts that $P(X=+1)=P(X=0)=P(X=-1)=\frac 13$. On the other hand, if we
assume that the absence of the spin-singlet state for deuteron ions means
that the Bose-Einstein triplet state is composed of two independently 
distributed spin $\frac 12$ particles then the model proposed in this paper
predicts $P(X=+1)=P(X=-1)=\frac 14$ and
$P(X=0)=\frac 12$. This should be testable by passing a beam of neutral 
deuteron atoms (not molecules) through a Stern-Garlach apparatus.
\footnote{Strictly speaking the failure of particles to form a spin-singlet 
state does not necessarily mean that the subsequent spin values of the triplet 
state are governed by the laws of independent probability. It may mean that
there is some type of dependent but non-deterministic relationship between 
the particles ($<1$).
This further highlights the importance of performing an experiment 
like that described above. If decoupled spin states imply statistical
independence then classification procedures become very simple. On the other
hand, if statistical independence fails to be observed then
the Bose-Einstein type statistic would have to be further sub-classified.}

\section {Conclusion:} 

In this paper a ``spin-coupling principle'' is derived which suggests a 
statistical classification of particles in terms of ISC states (spin-entangled
pairs) and non ISC states.
These ISC states appear to unify our understanding of atomic
orbitals, covalent bonding, paramagnetism, superconductivity, baryonic 
structure and so on. In summary, subatomic particles seem to form 
entangled {\bf pairs} whenever they are free to do so and there appears to be a 
universal principle at work, although the mechanism behind this coupling
would need to be further investigated.

Secondly, in contrast to the current paper, Pauli's version of the 
spin-statistics theorem imposes many other conditions on his particle system
including Minkowski
invariance, locality ( ``measurements at two space points with a space-like 
distance can never disturb each other''(\cite{pauli} p 721)), charge and 
energy densities.
However, the imposition of such extra conditions would seem to 
be unnecessary in the light of our current understanding of entanglement.
 
Finally, note that a connection between Bell's inequality{\cite Bell}
and rotational invariance has been established.

\section {Appendix}

In this section, an alternate proof of Lemma 2 is presented using a
probability argument. 
As before, let $s_n=s_n(\theta)$ represent the
spin of particle $n$ in the direction $\theta$, although the $\theta$ will
be dropped when there is no ambiguity. However, its presence should always be
understood. Also, let $\lambda_n=(q_n, s_n)$
represent the
quantum coordinates of particle $n$, with $s_n$ referring to the
spin coordinate in the direction $\theta$ and $q_n$ representing
occasionally, in the interest of clarity,
all other coordinates. In practice, 
$\lambda_n=(q_n, s_n)$ will represent the eigenvalues of an
operator defined on the Hilbert space ${\cal S}=L^2({\cal R}^3)\otimes
H_n$, where $H_n$ represents a two-dimensional spin space of the
particle $n$.  We will mainly work with $\lambda_n$ but sometimes,
there will be need to distinguish the $q_n$ from the $s_n$. This
distinction will also allow us to write the ket
$|\psi(\lambda)>=|\psi(q)>s$ where $s$ represents the spinor.

\begin{lem} Let 
\begin{eqnarray*}
|\psi(\lambda_1,\lambda_2)>
&=&c_1|\psi_1(\lambda_1)> \otimes
|\psi_2(\lambda_2)>+c_2|\psi_1(\lambda_2)> \otimes
|\psi_2(\lambda_1)>,\end{eqnarray*} where $c_1, c_2$ are independent of
$\lambda_1$, $\lambda_2$, represent an indistinguisable two particle
system defined on the space ${\cal S}_1\otimes {\cal S}_2$.
If ISC states for a system of two indistinguisable
and non-interacting particles occur only in the same q-orbital  
then the system
of particles can be represented by the Fermi-Dirac statistics.\end{lem}
{\bf Proof:} $q$-orbital spin-singlet states implies that 
$P(\lambda_1=\lambda_2)\le P(s_1=s_2)=0$. 
Therefore,
$<\psi(\lambda_1,
\lambda_1)|\psi(\lambda_1, \lambda_1)>=0$ and hence
$|\psi(\lambda_1, \lambda_1)>=0$, from the inner product
properties of a Hilbert space.
It follows, that $c_1=-c_2$ when the particles are
coupled and normalizing the wave function gives
$|c_1|=\frac{1}{\sqrt{2}}$. The result follows. 
QED 

A slightly less restrictive form of the lemma can also be proven using
essentially the same probability argument, as above.
\begin{lem} Let $|\psi(\lambda_1, \lambda_2)>$ denote a two 
particle state where the
$\lambda_1$ and $\lambda_2$ are as above. If the particles are in a 
spin-singlet state and non-interacting then their joint state can be expressed
in the form:
$$|\psi(\lambda_1,\lambda_2)>=\frac{1}{\sqrt 2}[|\psi(\lambda_1)|\psi(\lambda_2)
-|\psi(\lambda_2)>|\psi(\lambda_1)>].$$
In other words, coupled particles obey Fermi-Dirac statistics.
\end{lem}
{\bf Proof:} The particles are in a pure state of the subspace
$H_1\otimes H_2$ for non interactive and indistinguisable
particles. This implies that the general form of the two particle
state is given by:
$$|\psi(\lambda_1,\lambda_2)>=c_1|\psi_1(\lambda_1)> \otimes
|\psi_2(\lambda_2)>+c_2|\psi_1(\lambda_2)> \otimes
|\psi_2(\lambda_1)>.$$
Moreover, the particles are in a spin-singlet state and hence
$P(q_1, s_1;q_2, s_2)\le P(s_1=s_2)=0$. 
Therefore,
$<\psi(q_1,s_1;q_2,s_1)|\psi(q_1,s_1;q_2,s_1>=0$
and hence
$|\psi(q_1, s_1;q_1, s_2)>=0$, from the inner product
properties of a Hilbert space.
It follows, that $c_1=-c_2$ when the particles are
coupled and normalizing the wave function gives
$|c_1|=\frac{1}{\sqrt{2}}$. The result follows. 
QED 
\vspace{3mm}


\end{document}